\documentclass[letterpaper]{article} 
\usepackage{aaai2026}  
\usepackage{times}  
\usepackage{helvet}  
\usepackage{courier}  
\usepackage[hyphens]{url}  
\usepackage{graphicx} 
\usepackage{natbib}  
\usepackage{caption} 
\usepackage{algorithm}
\usepackage{algorithmic}

\usepackage{newfloat}
\usepackage{listings}
\DeclareCaptionStyle{ruled}{labelfont=normalfont,labelsep=colon,strut=off} 
\floatstyle{ruled}
\newfloat{listing}{tb}{lst}{}
\floatname{listing}{Listing}
\usepackage{tcolorbox}
\usepackage{amsmath}
\usepackage{multirow}
\usepackage{diagbox}

\title{SODBench: A Large Language Model Approach to Documenting Spreadsheet Operations}
\author {
    Amila Indika\textsuperscript{\rm 1},
    Igor Molybog\textsuperscript{\rm 2},
}
\affiliations {
    \textsuperscript{\rm 1}Department of Information and Computer Sciences, University of Hawaii at Manoa\\
    \textsuperscript{\rm 2}Department of Electrical and Computer Engineering, University of Hawaii at Manoa\\
    amilaind@hawaii.edu, 
    molybog@hawaii.edu
}

\usepackage{bibentry}

\begin{document}

\maketitle

\begin{abstract}
Numerous knowledge workers utilize spreadsheets in business, accounting, and finance. However, a lack of systematic documentation methods for spreadsheets hinders automation, collaboration, and knowledge transfer, which risks the loss of crucial institutional knowledge. This paper introduces Spreadsheet Operations Documentation (SOD), an AI task that involves generating human-readable explanations from spreadsheet operations. Many previous studies have utilized Large Language Models (LLMs) for generating spreadsheet manipulation code; however, translating that code into natural language for SOD is a less-explored area. To address this, we present a benchmark of 111 spreadsheet manipulation code snippets, each paired with a corresponding natural language summary. We evaluate five LLMs, GPT-4o, GPT-4o-mini, LLaMA-3.3-70B, Mixtral-8x7B, and Gemma2-9B, using BLEU, GLEU, ROUGE-L, and METEOR metrics. Our findings suggest that LLMs can generate accurate spreadsheet documentation, making SOD a feasible prerequisite step toward enhancing reproducibility, maintainability, and collaborative workflows in spreadsheets, although there are challenges that need to be addressed.
\end{abstract}

\begin{links}
    \link{Code}{https://github.com/AmilaIndika789/SheetCopilot}
    \link{Datasets}{https://figshare.com/s/1478ca752907477c4e4d}
\end{links}

\section{Introduction}
\subsection{Spreadsheet operations documentation}
Automated code documentation is a well-established challenge in software engineering, which mitigates developer burden and improves code maintainability~\cite{khan2022automatic}. For instance, recent research used Large Language Models (LLMs) to generate documentation for Python~\cite{poudel2024documint}. Similarly, in finance, accounting, and business, spreadsheets remain the primary tool for data processing~\cite{hesse2009electronic}, but spreadsheet operations pose unique documentation issues compared to codebases. Spreadsheet actions lack self-documentation, complicating the reconstruction of action sequences and the underlying rationale, even when the spreadsheets are well-organized. Unlike software development, spreadsheet workflows lack collaborative tools such as GitHub, which provide version control and structured documentation. Often, many users conduct manual documentation and version control~\cite{roy2017spreadsheet}. These poor practices lead to significant issues, particularly during employee turnover, as undocumented institutional knowledge is lost, and onboarding new staff can take more than six months due to insufficient documentation.
In regulated industries, such as finance, poor documentation increases non-compliance risks due to the need for an auditable trail. Lack of knowledge transfer can lead to mistakes, replication, and collaboration challenges without systematic documentation. Hence, automated Spreadsheet Operations Documentation (SOD) solutions are crucial for overcoming these challenges. The lack of datasets and tools creates a bottleneck for the SOD problem that this work seeks to address.

\subsection{State-of-the-art in spreadsheet manipulation}
Despite spreadsheet ubiquity, they are prone to errors~\cite{panko1998we, panko2016we}. Common issues such as formula errors, incorrect data entry, and inconsistent logic can lead to significant financial losses, flawed decision-making, and misinterpretation of data~\cite{powell2008critical}. As spreadsheets increase in size and complexity, they require extensive data cleaning, transformation, and manipulation, which becomes challenging due to varying user expertise.

Traditional methods such as Visual Basic for Applications (VBA) for Excel~\cite{walkenbach2013excel} and Google Apps Script (GAS)~\cite{ganapathy2016learning}, enable manipulation but require programming skills, posing a barrier for many users. Similarly, using Python for spreadsheet manipulation~\cite{zumstein2021python} offers powerful functionality but requires programming. Business intelligence platforms, such as Power BI and Tableau, offer advanced data manipulation features but are costly, require substantial training, making them less ideal for routine spreadsheet tasks~\cite{negash2008business}. Rule-based manipulation approaches~\cite{shigarov2017rule, shigarov2019tabbyxl} are useful for simple, pattern-driven operations but lack the flexibility to handle complex, multi-step decisions. Consequently, there is a need for effective solutions that simplify advanced spreadsheet manipulation and enhance transparency and reproducibility.

\subsection{LLM-driven spreadsheet manipulation}
Latest LLM developments enhance their language comprehension, code generation, and multi-step task performance abilities, making them effective for manipulating structured data in spreadsheets.~\cite{naveed2023comprehensive, jiang2024survey, li2025fundamental, lu2025large}. LLMs can interpret user intent expressed in plain language and translate it into actionable tasks, such as generating formulas, identifying patterns, and executing multi-step transformations, without requiring user intervention~\cite{chen2024sheetagent, li2024sheetcopilot}.

An advantage of LLM-driven spreadsheet manipulation is increased accessibility. For example, business users can describe operations in natural language (e.g., "Calculate average monthly sales by product category"), and the LLM can generate the formulas. This language interface simplifies tasks for non-programmers, reducing human errors by automating tedious tasks. For example,  LLMs support advanced dataset merging and pivot table creation, requiring minimal user setup. They also make complex spreadsheet features more accessible to users of all skill levels. However, capturing user actions in spreadsheets remains a challenge, which is crucial for effective spreadsheet manipulation.

\subsection{Goals and research questions}
\begin{figure*}[!htbp]
    \centering
    \includegraphics[width=0.75\linewidth]{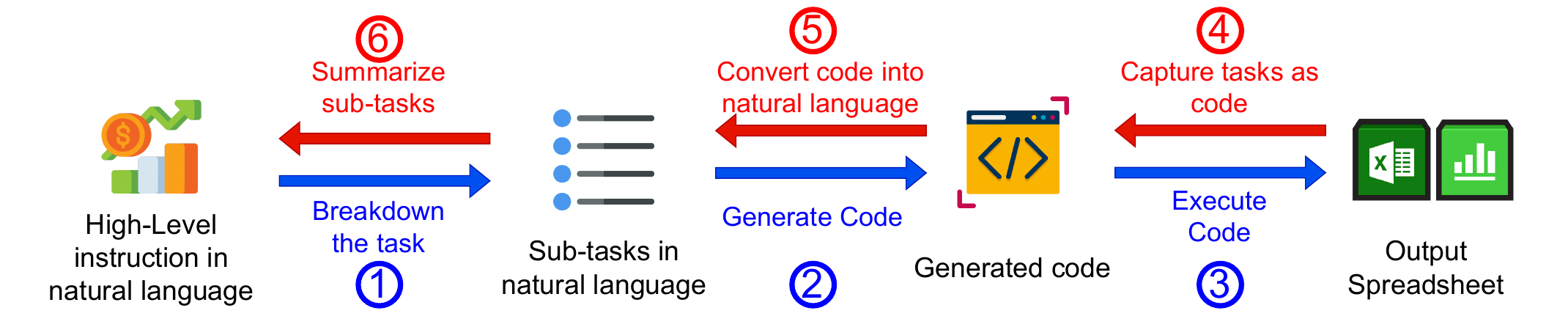}
    \caption{Overview of spreadsheet workflows. The blue arrows indicate the traditional approach where spreadsheet agents break down user instructions into subtasks and create executable code. In contrast, our work emphasizes the reverse process (red arrows), translating executed actions back into natural language to document and explain user operations.}
    \label{fig: reverse_process_overview}
\end{figure*}

Many LLM-based tools can manipulate spreadsheets by generating formulas and automating tasks, but a key challenge is capturing and translating user actions into natural language descriptions. Key aspects, such as reproducibility, transparency, knowledge transfer, collaboration, automation, and interpretability, are limited without this capability. This problem is severe in organizations where users with different technical skills share spreadsheets.

We motivate this problem through the following use case: \textit{A senior analyst completes complex data manipulations in a spreadsheet and shares it with an intern, who must replicate the workflow without access to the analyst's intent or reasoning}. This straightforward but practical scenario highlights three distinct yet interconnected challenges:

\begin{enumerate}
\item \textbf{Reproducibility and transparency:}
The intern needs the analyst's actions to replicate their work. By translating the analyst's actions into plain language, we can automatically generate documentation that provides an audit trail, enabling reproducibility without manual effort.

\item \textbf{Knowledge transfer and collaboration:}  
The intern may lack awareness of advanced technical skills and spreadsheet features. Using auto-generated descriptions of the analyst's actions enables effortless knowledge transfer.

\item \textbf{Automation, evaluation, and reasoning of AI agents:}  
Recording the analyst's actions creates valuable data for developing AI agents. We can record the analyst's actions and convert them into plain language to train LLMs for reasoning about GUI manipulation. This approach enhances LLMs' reasoning in structured and GUI environments like spreadsheets, beyond just text and code.

\end{enumerate}

Our research question aims to formalize this gap in light of these goals and explore how to utilize LLMs to bridge it.

\begin{tcolorbox}[
  title=Research Question (RQ),
  colback=gray!20,
  colframe=gray!70,
  coltitle=black,
  colbacktitle=gray!40,
  fonttitle=\bfseries,
  sharp corners=south,
  boxrule=0.8pt,
  width=0.48\textwidth,
  boxsep=0.6pt,
]
How can LLMs translate actions in spreadsheets into natural language to (1) enable reproducibility and transparency for audits, (2) facilitate knowledge transfer and collaboration between users with varying expertise, and (3) generate reasoning traces for training and evaluating AI agents in GUI environments?
\end{tcolorbox}

Screen recording is a straightforward way to capture spreadsheet interactions, but manually analyzing recorded videos or using external tools has drawbacks: it may miss semantic details, cannot generalize tasks, is time-consuming, has resource limitations, and requires post-processing.

Instead, we propose capturing spreadsheet workflows as code and translating them into natural language for structured representation of user actions (see red arrows in Figure~\ref{fig: reverse_process_overview}). Specifically, when a user completes a task, we can capture it as code (e.g., via macro recording), decompose it into natural language subtasks, and synthesize high-level instructions. This reverse approach provides interpretable, human-readable descriptions of spreadsheet workflows.

Thus, the research problem of translating low-level spreadsheet operations into natural language to support reproducibility, knowledge transfer, and interpretable automation aligns with our central research question.

To this end, we introduce a benchmark dataset mapping spreadsheet manipulation code into natural language, enabling AI agents to enhance documentation by converting spreadsheet operations into natural language.

\subsection{Contributions}
For clarity, here are our main contributions from this paper:
\begin{enumerate}
    \item We formalize the reverse spreadsheet manipulation problem, Spreadsheet Operations Documentation (SOD), which involves generating natural language descriptions for spreadsheet operations.
    \item We created a novel benchmark dataset\footnote{https://figshare.com/s/1478ca752907477c4e4d}, SODBench, with 111 validated spreadsheet manipulation task instances, each paired with xwAPI code and natural language.
    \item We evaluate the code-to-natural language translation performance of five LLMs using multiple NLP metrics (BLEU, GLEU, ROUGE-L, METEOR) and analyze statistical significance using confidence intervals and t-tests.
    \item We present a Retrieval-Augmented Generation (RAG) pipeline to generate JavaScript (JS) code for spreadsheet automation\footnote{https://github.com/AmilaIndika789/SheetCopilot}, demonstrating the viability of combining LLMs with external API documentation for automation.
    \item We outline how the reverse spreadsheet manipulation approach could be extended into a practical tool, such as a browser extension in Google Spreadsheets or a plugin in Excel, for automated SOD in real-world workflows.
\end{enumerate}

\section{Related Work}
\label{section: related_work}
LLM researchers have developed numerous benchmarks for evaluating their spreadsheet manipulation skills. A notable example is SheetCopilot~\cite{li2024sheetcopilot}, which includes a spreadsheet agent and a benchmark of 221 tasks to evaluate LLMs' control over spreadsheet operations. SheetCopilot offers a framework of atomic actions that enhances spreadsheet functionalities. It converts high-level natural language instructions into atomic actions, translates them into Python code, and executes the code on the spreadsheet to achieve the desired outcome. FLAME~\cite{joshi2024flame} utilizes a transformer-based model for spreadsheet formula manipulation, outperforming state-of-the-art (SOTA) models such as CodeT5 and Codex in formula repair, completion, and retrieval tasks. FLAME facilitates spreadsheet manipulation using LLMs, but focuses primarily on formula operations.

\citet{chen2024sheetagent} introduces an LLM agent evaluation benchmark on sequential-reasoning, long-horizon planning, and multiple category tasks. This benchmark comprises dependent tasks that require discrete actions, enabling automatic evaluation using various metrics. The authors also introduce SheetAgent~\cite{chen2024sheetagent}, an LLM-based agent designed for improved reasoning for spreadsheet manipulation. SpreadsheetCoder~\cite{chen2021spreadsheetcoder} explores the synthesis of spreadsheet formulas. SpreadsheetCoder incorporates contextual information, such as table headers, along with a BERT-based model to synthesize formulas. However, this approach is limited to formula generation.

SpreadsheetLLM~\cite{tian2024spreadsheetllm} presents a method for encoding spreadsheets into a compressed format for efficient processing by LLMs, along with a ``Chain of Spreadsheets" approach to improve analysis in downstream applications. However, SpreadsheetLLM lacks cell formatting and semantic information in its compressed representations.

Our research builds upon prior work in spreadsheet manipulation, focusing on SOD. Similar to SheetCopilot and SheetAgent, we integrate LLMs with spreadsheets, but we emphasize generating descriptive documentation of performed operations rather than sequential reasoning over tasks. Unlike FLAME and SpreadsheetCoder, which focus on formula manipulation, our work addresses a broader range of spreadsheet actions, including pivot tables, chart creation, formatting, and multi-step edits. SpreadsheetLLM enhances spreadsheet comprehension through compression for machine tasks, while we focused on improving human interpretability by providing an audit trail of user actions.

Furthermore, inspired by benchmarks like SheetCopilot's 221-task benchmark and SheetAgent's evaluation framework, we introduce SODBench. SODBench evaluates LLMs' ability to map code into interpretable SOD.

\textbf{Summary:} \textit{ While prior work advances automation and control in spreadsheet manipulation, our research complements this by making those manipulations transparent, interpretable, and communicable through spreadsheet documentation, which supports reproducibility, collaboration, and knowledge transfer in spreadsheet workflows.}

\section{Methods}
\label{section: methods}
\subsection{Problem statement}
Figure~\ref{fig: reverse_process_overview} illustrates spreadsheet manipulation using the forward approach (blue arrows: 1, 2, 3) and SOD with the backward approach (red arrows: 4, 5, 6). This research focuses on SOD in capturing and documenting user actions.

\subsection{Proposed method}
Our primary focus is step 5 in Figure~\ref{fig: reverse_process_overview}, which translates spreadsheet manipulation code into natural language. We tackle this through two components: \textit{(1) creating a benchmark dataset of code-summary pairs and (2) evaluating LLM performance on this task}. We discuss step 4, which involves capturing user actions from spreadsheets, reviewing traditional methods for action extraction, and adapting them for our benchmarking pipeline. In step 6, mapping natural language subtasks to higher-level instructions, is essentially text summarization, and extensive research has been conducted on the topic. Hence, we do not further explore step 6 in this study. 

\subsection{Translation of code to plain language (step 5)}
\subsubsection{Generate a dataset mapping code and natural language}
\label{subsection: dataset_generation}
We require a dataset that connects spreadsheet manipulation code to natural language descriptions. Since none exists, we create a synthetic benchmark by generating code and description pairs, serving two purposes. It helps to \textit{1) evaluate translation of spreadsheet manipulation code into natural language} and 2) \textit{provide a benchmark dataset for research}.

\noindent\underline{\textit{Data sources}}: We create our dataset by reverse-engineering SheetCopilot's implementation~\cite{copilot_implementation}, making adjustments to meet our research goals. Our analysis is based on their 221 benchmark task instances~\cite{li2024sheetcopilot}.

\noindent\underline{\textit{Data generation process}}: 
The data generation objective is to generate code and corresponding natural language instructions for user actions when solving a spreadsheet task instance. We used 221 benchmark task instances from SheetCopilot to create a dataset of code-to-natural language pairs. 

\noindent\underline{\textit{SheetCopilot's benchmark dataset structure}}: SheetCopilot organizes 221 benchmark task instances into five broad categories of atomic actions, ``Entry and manipulation", ``Management", ``Formatting", ``Chart", and ``Pivot table" categories. Some task instances span multiple categories, such as ``Chart and Pivot table" or ``Management and Chart". The dataset comprises 28 Excel workbooks (see Appendix~A). SheetCopilot uses \emph{xwAPI}, a set of atomic operations, for generalized Excel spreadsheet manipulation.

\begin{figure}[!hbtp]
    \centering
    \includegraphics[width=0.7\linewidth]{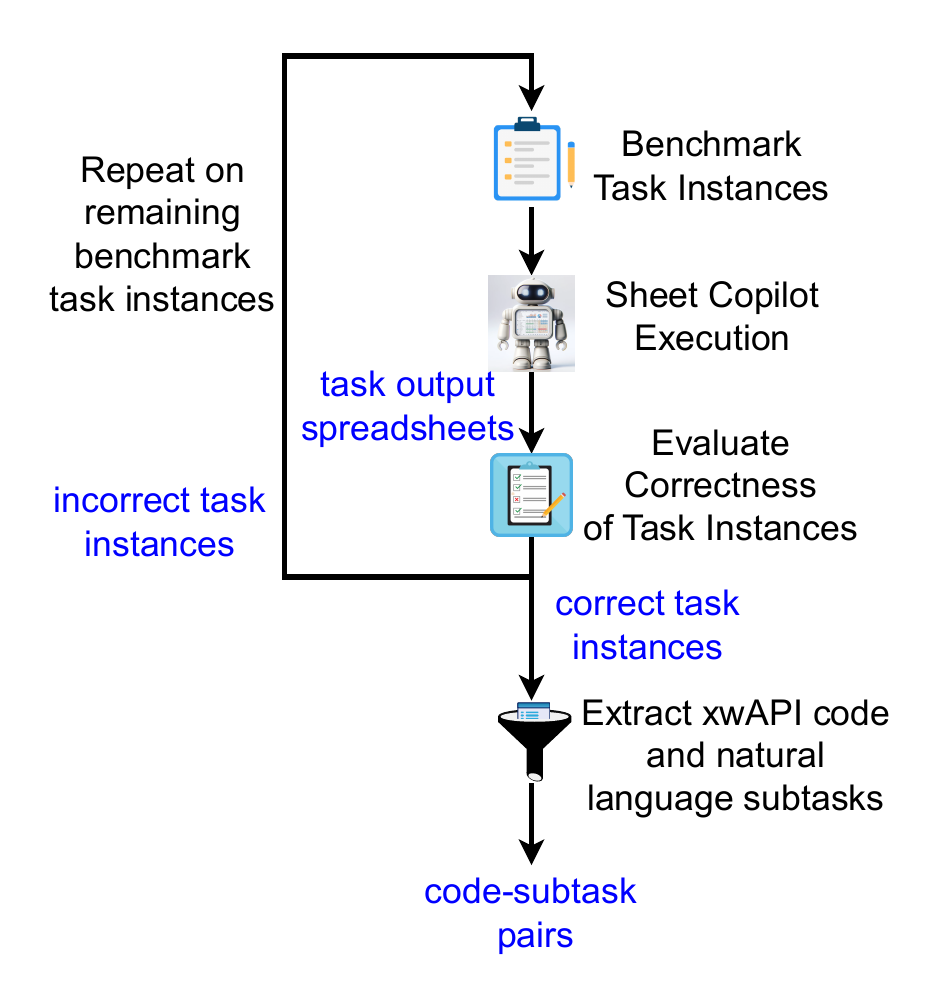}
    \caption{Data generation workflow: SheetCopilot performs 221 benchmark tasks to create output spreadsheets. xwAPI code and natural-language description pairs are extracted from correct instances, while incorrect ones are re-executed.}
    \label{fig: data_generation}
\end{figure}

\noindent\underline{\textit{Data Generation Workflow}}: Figure~\ref{fig: data_generation} shows our data generation workflow. 

This process corresponds to step 2 in Figure~\ref{fig: reverse_process_overview}. We opted for an automated approach to build the benchmark for better scalability instead of manual coding. We chose GPT-4o-mini for data generation because of its compatibility with SheetCopilot, cost-accuracy balance, and budget constraints. We used GPT-4o when GPT-4o-mini plateaued.

We evaluated each task instance's output against ground-truth Excel outputs from the SheetCopilot dataset using a validation script provided by SheetCopilot. The validation script verifies the accuracy of each benchmark task by comparing its output to the corresponding ground-truth Excel workbook, by iterating over all worksheets to ensure correctness. The script compares cell data column-wise, evaluates charts, examines pivot tables, verifies filters, and checks formatting for each worksheet. It then assesses the alignment of frozen panes in each workbook.

We extracted the xwAPI code and corresponding natural language subtasks for correct task instances. These pairs create our core dataset, which represents successful translations from code to natural language. We recycled incorrect task instances from the benchmark to iteratively expand the dataset, gradually accumulating accurate code-subtask pairs.

We manually tuned LLM hyperparameters to address frequent data generation failures, changing \textit{temperature [0.4, 1.2]}, setting \textit{max\_total\_tokens to 16,384}, \textit{max\_new\_tokens to 256}, and \textit{API timeout to 50 seconds}. This process yielded 111 valid task instances from the original 221.

\noindent\underline{\textit{Generated dataset}}: The final dataset comprises 111 valid code-to-natural language mappings, each pairing an \emph{xwAPI} code block with a corresponding human-readable subtask description. We used this \emph{xwAPI} code as input for our LLM comparison pipeline.

\subsubsection{Evaluating LLMs' ability on the translation task}


\noindent\underline{\textit{Comparing LLM performance}}: We evaluate how well LLMs translate \emph{xwAPI} code into natural language subtasks, measuring accuracy against ground-truth representations using Hugging Face's text evaluation metrics~\cite{huggingface_evaluate_metrics}. We used BLEU~\cite{papineni2002bleu}, GLEU~\cite{wu2016google}, ROUGE-L~\cite{lin2004rouge}, and METEOR~\cite{banerjee2005meteor} metrics to measure the similarity between the generated and reference subtasks' plain language representations.

\noindent\underline{\textit{Experimental Setup}}: We deployed a virtual machine running Windows 11 with Microsoft Excel 2021 installed. We customized the SheetCopilot repository by accessing proprietary LLMs, GPT-4o and GPT-4o-mini via the OpenAI API, and open-source LLMs Llama-3.3-70B, Mixtral-8-7B, and Gemma2-9B via the Groq API.

LLM pretraining is task-agnostic, but adapting to new tasks often faces challenges due to the limited availability of domain-specific training data, requiring fine-tuning~\cite{brown2020language}. Few-shot learning~\cite{gao2020making, brown2020language} and prompt engineering~\cite{reynolds2021prompt, sahoo2024systematic} are standard techniques to mitigate this issue. Prior studies indicate that LLM performance plateaus beyond a certain number of provided examples~\cite{zhao2021calibrate, perez2021true}. To validate this, we conducted an auxiliary experiment varying the number of few-shot examples and evaluated the resulting code-to-natural language translation performance using established NLP metrics. Our findings confirmed that performance gains plateaued beyond a 4-shot configuration. Thus, we adopted 4-shot learning for all models except Mixtral-8-7B, which required fewer shots due to its limited context window. We provided further details in Appendix B.

We explored prompt engineering by using xwAPI documentation, 4-shot examples, and structured templates from GPT Prompt Crafter~\cite{prompt_crafter, open_ai_prompt_engineering}. We also provided a representative prompt in Appendix~C. In our 4-shot configuration, we evaluated each LLM's performance on 107 out of 111 valid task instances using the evaluation metrics, maintaining a \textit{temperature} of 0.5 for all evaluations. 

\subsection{Capturing spreadsheet actions (step 4)}
\begin{figure*}[!htbp]
    \centering
    \includegraphics[width=0.8\linewidth]{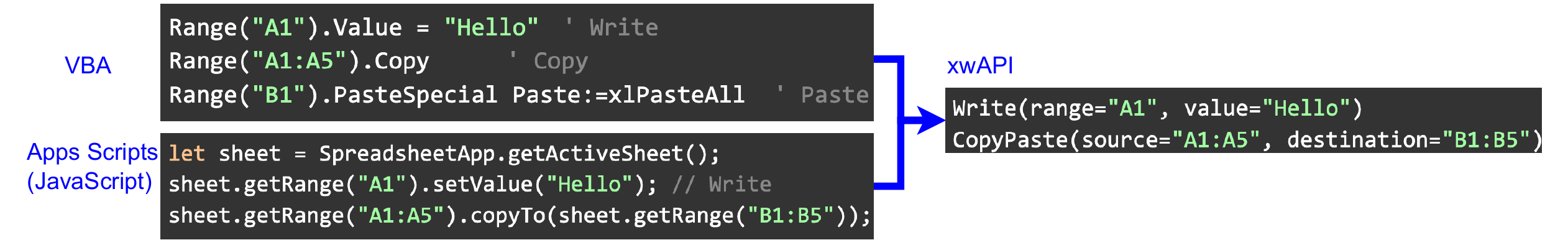}
    \caption{Example of atomic-action translation to xwAPI: user actions—writing to cell A1 and copying A1:A5 to B1:B5—are shown in VBA (top-left) and Google Apps Script/JS (bottom-left), then mapped to xwAPI equivalents (right). Each atomic operation (Write, CopyPaste) is preserved, demonstrating code conversion into xwAPI through simple syntax substitution.}
    \label{fig: code_translation}
\end{figure*}

We must highlight that we cannot natively capture users' actions on spreadsheets in the \emph{xwAPI} format. However, several methods exist for recording user actions. These include: \textit{1) VBA macros in desktop Excel}, \textit{2) Office Script via the Action Recorder in web Excel}~\cite{office_scripts}, and \textit{3) Google Apps Script in Google Sheets}. Furthermore, the Excel JavaScript API~\cite{excel_js_api} facilitates the interaction with Excel spreadsheets. After recording user actions in code, translating VBA or Apps Scripts into \emph{xwAPI} involves mapping each atomic action to its equivalent syntax in the target language. For example, atomic actions, such as writing to a cell or copying ranges, can be easily translated into \emph{xwAPI} by adjusting the syntax, as shown in Figure~\ref{fig: code_translation}. While we demonstrate that translating VBA or JS to \emph{xwAPI} is feasible, we also show that this intermediate mapping can be bypassed in our supplementary experiments (see Appendix~D). Instead of using \emph{xwAPI} code, we show that we can generate JS code generated by an LLM with Retrieval-Augmented Generation (RAG) through LangChain. Then, we can even skip the additional translation step from VBA or JS into \emph{xwAPI}. 

\section{Results}
\label{section: results}

\begin{table*}[!htbp]
\centering
\caption{Mean evaluation metrics ($\pm$MOE with 95\% CI) for the LLMs. Gemma-2×9 B is significantly the worst, Mixtral-8×7 B is the next lowest, and GPT-4o-mini, Llama-3.3-70B, and GPT-4o are statistically not comparable with overlapped MOEs.}
\label{tab: llm_evaluation_metrics}
\begingroup
\def\arraystretch{1}
\begin{tabular}{|c|c|c|c|c|}
\hline
\textbf{LLM} & \textbf{mean BLEU $\uparrow$ } & \textbf{mean GLEU $\uparrow$ } & \textbf{mean ROUGE-L $\uparrow$ } & \textbf{mean METEOR $\uparrow$ } \\ \hline
gemma2 9B     & 0.042 $\pm$ 0.018 & 0.060 $\pm$ 0.020 & 0.155 $\pm$ 0.043 & 0.088 $\pm$ 0.032 \\ \hline
llama-3.3-70B & 0.257 $\pm$ 0.034 & 0.322 $\pm$ 0.029 & 0.524 $\pm$ 0.032 & 0.464 $\pm$ 0.034 \\ \hline
mixtral-8x7B  & 0.080 $\pm$ 0.018 & 0.114 $\pm$ 0.017 & 0.260 $\pm$ 0.025 & 0.331 $\pm$ 0.024 \\ \hline
gpt-4o       & 0.254 $\pm$ 0.034 & 0.317 $\pm$ 0.028 & 0.516 $\pm$ 0.029 & 0.494 $\pm$ 0.035 \\ \hline
gpt-4o-mini  & 0.265 $\pm$ 0.036 & 0.327 $\pm$ 0.031 & 0.532 $\pm$ 0.032 & 0.498 $\pm$ 0.036 \\ \hline
\end{tabular}
\endgroup
\end{table*}


Table \ref{tab: llm_evaluation_metrics} shows evaluation results across five LLMs, with means calculated for each metric using Equation \ref{equation: mean_metric}.

\begin{tcolorbox}[
  colframe=black,
  colback=white,
  boxrule=1pt,
  arc=3mm,
  left=1pt,
  right=1pt,
  top=1pt,
  bottom=1pt,
  boxsep=0.5pt,
]
    \begin{equation}
        \label{equation: mean_metric}
        \text{Mean metric} = \frac{1}{N} \sum_{i=1}^{N} (\text{metric})_{i} \;\;\;
        \text{where } N = 107
    \end{equation}
    \begin{equation*}
        \text{metric} \in \text{\{BLEU, GLEU, ROUGE-L, METEOR\}}
    \end{equation*}
\end{tcolorbox}

The margin of error (MOE) at a 95\% Confidence Interval (CI) is calculated using Equation \ref{equation: margin_of_error}, which reflects statistical variability in model performance.

\begin{tcolorbox}[
  colframe=black,
  colback=white,
  boxrule=1pt,
  arc=3mm,
  left=1pt,
  right=1pt,
  top=1pt,
  bottom=1pt,
  boxsep=0.5pt,
]
    \begin{equation}
        \label{equation: margin_of_error}
        \text{Margin of Error (MOE)} = Z_{95} \times \frac{\sigma}{\sqrt{N}}
    \end{equation}
    \begin{equation*}
        \text{where } N = 107,\ \sigma = \text{Standard deviation},
    \end{equation*}
    \begin{equation*}
        Z_{95} = \text{Z-score for a 95\% Confidence Interval (CI)}
    \end{equation*}
\end{tcolorbox}

Table \ref{tab: llm_evaluation_metrics} shows that Gemma2-9B has the lowest mean across all metrics, with non-overlapping CIs with LLMs. This non-overlap in the MOE statistically confirms that Gemma2-9B significantly underperforms compared to other LLMs. Similarly, Mixtral-8x7B displays the second-lowest performance across metrics, with non-overlapping CIs reinforcing its statistically inferior standing relative to others.

\begin{tcolorbox}[
  colframe=black,
  colback=white,
  boxrule=1pt,
  arc=3mm,
  left=1pt,
  right=1pt,
  top=1pt,
  bottom=1pt,
  boxsep=0.5pt,
]
    \begin{equation}
        \label{equation: t_test_sd}
        t = \frac{\bar{x}_1 - \bar{x}_2}{\sqrt{\frac{\sigma_1^2}{N_1} + \frac{\sigma_2^2}{N_2}}} \;\;\;
        \text{where } N_1 = N_2 = 107
    \end{equation}
    \begin{equation*}
        \bar{x}_1, \bar{x}_2 \text{ are the sample means}
    \end{equation*}
    \begin{equation*}
        \sigma_1, \sigma_2 \text{ are the sample standard deviations}
    \end{equation*}
\end{tcolorbox}

In Table~\ref{tab: llm_evaluation_metrics}, the point estimates suggest that GPT-4o-mini has the highest mean scores, followed by Llama-3.3-70B and GPT-4o. However, this ranking is misleading due to overlapping MOEs. The intersecting 95\% CIs show that the result is not statistically significant to claim that any model outperforms the others based solely on the point estimates. We conducted a pairwise two-sample t-test to rigorously evaluate whether performance differences among these three models are statistically significant. We calculated t-values using Equation~\ref{equation: t_test_sd} and p-values through a two-tailed test under a standard normal distribution, as shown in Equation~\ref{equation: p_value}.

\begin{tcolorbox}[
  colframe=black,
  colback=white,
  boxrule=1pt,
  arc=3mm,
  left=0.5pt,
  right=0.5pt,
  top=0.5pt,
  bottom=0.5pt,
  boxsep=0.5pt,
]
    \begin{equation}
        \label{equation: p_value}
        p = 2 \cdot \left(1 - \Phi\left(|t|\right)\right)
    \end{equation}
    \begin{equation*}
        \text{where } \Phi \text{ is CDF of normal distribution}
    \end{equation*}
\end{tcolorbox}



Our two-sample t-test shows all p-values exceed 0.05 (see Appendix~E), indicating that the performance differences among the top three LLMs are not statistically significant. Thus, these models are comparable in performance for SOD.

Although prior research shows LLM performance typically follows a power-law relationship with model parameters, training data size, and compute budget~\cite{kaplan2020scaling}, larger LLMs do not necessarily ensure better performance on specialized tasks~\cite{hoffmann2022training}. Despite Mixtral-8x7B and Gemma-2x9B having more parameters than GPT-4o-mini, the latter outperforms them in all metrics, which demonstrates that model size alone does not determine performance, particularly in tasks such as SOD.

Gemma2-9B's poor performance is likely due to its architectural and pre-training limitations, possibly stemming from a limited training corpus, which weakens its ability to handle contextualized spreadsheet tasks. Mixtral-8x7B's underperformance is likely due to its narrower context window and using a 1-shot compared to others that used a 4-shot context. These limitations may have hindered its SOD ability.

Prior work highlights the Mixtral models' limited generalization capacity, even in low-complexity domains~\cite{yan2024efficient, dorfner2024open, madhusudhan2024llms}. Our findings suggest this stems from architectural design, inadequate training data, and restricted few-shot learning ability.

These findings highlight the importance of integrating multiple components for LLM success in specialized areas such as SOD. These include the model's architecture, training data quality and coverage, and techniques such as instruction tuning~\cite{wei2021finetuned}, prompt engineering~\cite{reynolds2021prompt}, few-shot configuration~\cite{gao2020making}, chain-of-thought reasoning~\cite{wei2022chain}, and domain-specific fine-tuning~\cite{brown2020language}. The interplay of these factors is more crucial than model scale in optimizing LLM performance for applications such as SOD~\cite{lu2025large}.

\section{Discussion}
\label{section: discussion}
Here are key takeaways from our findings.

\noindent\textbf{$\star$ SODBench is sufficiently challenging for LLMs.} \newline
Our results show that SOTA LLMs struggle to translate spreadsheet manipulation code into natural language with SODBench. The evaluation metrics yield mean scores in the range [0, 1], indicating a balanced dataset, neither saturated (close to 1) nor infeasible (close to 0). SODBench balances non-trivial, structured tasks that require nuanced reasoning and contextual understanding, which makes our dataset a robust resource for evaluating LLM capabilities in SOD.

\noindent\textbf{$\star$ Spreadsheet operations documentation is feasible.} \newline
Our experiments show that the top three LLMs achieve average scores of about 0.25, 0.32, 0.52, and 0.49 for BLEU, GLEU, ROUGE-L, and METEOR, respectively. These empirical results prove that LLMs can generate meaningful, semantically relevant summaries from spreadsheet manipulation code. With advanced fine-tuning, few-shot learning \& prompt engineering, we can enhance performance metrics. Hence, our study confirms effective code-to-natural language translation using LLMs, addressing our core RQ.

\noindent\textbf{$\star$ LLMs support reproducibility, transparency, collaboration, and automation in spreadsheets.} \newline
Our study shows that translating spreadsheet code into natural language is possible, improving the reproducibility, interpretability, shareability, and automation of workflows with LLMs. Although not formally tested, our findings suggest that LLMs support these broader goals. Our future research will focus on rigorously validating these aspects.

\noindent\textbf{$\star$ The performance of LLMs on SOD is influenced by multiple architectural and methodological factors.} \newline
We showed that larger LLMs do not always ensure better performance. For instance, Gemma-2-9B and Mixtral-7x8B, despite being large LLMs, underperformed in code-to-natural language translation. In contrast, models like GPT-4o, GPT-4o-mini, and LLaMA-3.3 showed better results, which highlights the significance of in-context learning, prompt engineering, domain-specific fine-tuning, and chain-of-thought reasoning in improved performance. Moreover, our performance rankings align with the ChatBot Arena~\cite{chiang2024chatbot, lmarena2025leaderboard}, indicating that GPT-4o, GPT-4o-mini, and LLaMA-3.3 outperform Gemma-2-9B \& Mixtral-7x8B. These results show that LLM performance on SOD tasks depends on multiple factors.

\noindent\textbf{$\star$ Evaluation metrics must reflect task objectives.} \newline
We used various evaluation metrics for natural language similarity, including BLEU, GLEU, ROUGE-L, and METEOR, which are commonly used in natural language generation and available via Hugging Face. We evaluated code correctness using Pass@1 and Exec@1 metrics, as defined by SheetCopilot~\cite{li2024sheetcopilot} and other code-focused LLM evaluations~\cite{chen2021evaluating}. These metrics assess LLM performance in natural language and code generation, and future research should adopt similar approaches.

\section{Limitations and Future Work}
\label{section: limitations_and_future_work}
Our SOD method is promising, but has limitations. First, SODBench includes only 111 code-to-natural language pairs, which may not cover all spreadsheet operations. Therefore, future work should expand the dataset. Second, we evaluated a limited set of LLMs and in-context learning setups due to budget and time constraints, impacting the generalizability of our findings. Therefore, future studies should explore more LLMs, few-shot learning setups, and improved prompt engineering to enhance performance and validate our approach. Third, we claim that LLMs enhance reproducibility, knowledge transfer, and automation in spreadsheets, but our study only lays the groundwork and does not rigorously prove these abilities. Hence, we encourage future research to evaluate LLMs' impact on reproducibility and automation in autonomous and human-in-the-loop settings.

\noindent\textbf{Deployment Path:} Our current RAG pipeline generates JS code for Excel manipulation as a proof of concept. Future development could lead to a browser extension (Google Sheets) or a plugin (Microsoft Excel) that captures user actions as executable JS code. These captured actions, along with relevant JS API context, would allow an LLM to create natural language summaries, reducing the SOD burden for users and simplifying the process to just a few clicks.

\section{Conclusion}
\label{section: conclusion}
We present a novel approach for SOD using LLMs. We present SODBench, a dataset of code-to-natural language pairs, and evaluate five LLMs, comparing their ability to generate accurate and interpretable descriptions. GPT-4o, GPT-4o-mini, and LLaMA-3.3 show promising results. Our findings show that effective SOD performance relies not just on LLM scale, but also on architecture, prompt engineering, and few-shot configurations. Our implementation can be enhanced as a deployable extension or plugin for automating SOD. We highlight how SOD improves reproducibility, knowledge transfer, and collaboration in spreadsheets. We introduce SODBench to encourage further exploration and invite future work to enhance the dataset, boost generation quality, and integrate LLMs into spreadsheet tools.

\bibliography{aaai2026}

\clearpage
\onecolumn
\section*{Reproducibility Checklist}
\begin{itemize}
    \item This paper:
    \begin{itemize}
        \item Includes a conceptual outline and/or pseudocode description of AI methods introduced: \textbf{Yes, Figure 1}.
        \item Clearly delineates statements that are opinions, hypotheses, and speculation from objective facts and results: \textbf{Yes, sections 4, 5, and 6}.
        \item Provides well-marked pedagogical references for less-familiar readers to gain the background necessary to replicate the paper: \textbf{Yes, subsections 1.2 \& 1.3 and section 2}.
    \end{itemize}
    \item Does this paper make theoretical contributions? \textbf{No, application-oriented}.
    \item Does this paper rely on one or more datasets? \textbf{Yes}.
    \begin{itemize}
        \item A motivation is given for why the experiments are conducted on the selected datasets: \textbf{Yes, subsection 3.3, line numbers 254 to 260}.
        \item All novel datasets introduced in this paper are included in a data appendix: \textbf{No data appendix, but the dataset is shared through an external link--Datasets link and footnote 1}.
        \item All novel datasets introduced in this paper will be made publicly available upon publication of the paper with a license that allows free usage for research purposes: \textbf{Yes, shared under arXiv.org perpetual, non-exclusive license}.
        \item All datasets drawn from the existing literature (potentially including authors' own previously published work) are accompanied by appropriate citations: \textbf{Yes, SheetCopilot~\cite{li2024sheetcopilot}}.
        \item All datasets drawn from the existing literature (potentially including authors' own previously published work) are publicly available: \textbf{Yes, SheetCopilot GitHub~\cite{copilot_implementation}}
        \item All datasets that are not publicly available are described in detail, with an explanation of why publicly available alternatives are not scientifically satisfactory: \textbf{NA}.
    \end{itemize}
    \item Does this paper include computational experiments? \textbf{Yes}.
    \begin{itemize}
        \item This paper states the number and range of values tried per (hyper-) parameter during development of the paper, along with the criterion used for selecting the final parameter setting: \textbf{Yes, line numbers 303 to 306, and 350 to 351}.
        \item Any code required for pre-processing data is included in the appendix: \textbf{No code appendix, but a GitHub repository with source code is shared}.
        \item All source code required for conducting and analyzing the experiments is included in a code appendix: \textbf{No code appendix, but a GitHub repository is made public}.
        \item All source code required for conducting and analyzing the experiments will be made publicly available upon publication of the paper with a license that allows free usage for research purposes: \textbf{Yes}.
        \item All source code implementing new methods has comments detailing the implementation, with references to the paper where each step comes from: \textbf{Partial, detailed comments are included for most of the new methods}.
        \item If an algorithm depends on randomness, then the method used for setting seeds is described in a way sufficient to allow replication of results: \textbf{NA, no explicit randomness except from LLMs}.
        \item This paper specifies the computing infrastructure used for running experiments (hardware and software), including GPU/CPU models, amount of memory, operating system, names and versions of relevant software libraries and frameworks: \textbf{Partial, line numbers 322 to 327}.
        \item This paper formally describes the evaluation metrics used and explains the motivation for choosing these metrics: \textbf{Yes, line numbers 314 to 321}.
        \item This paper states the number of algorithm runs used to compute each reported result: \textbf{No, we do not run the experiments repeatedly}.
        \item Analysis of experiments goes beyond single-dimensional summaries of performance (e.g., average, median) to include measures of variation, confidence, or other distributional information: \textbf{Yes, confidence level and margin of errors are included--Table 1}
        \item The significance of any improvement or decrease in performance is judged using appropriate statistical tests (e.g., Wilcoxon signed-rank): \textbf{Yes, pairwise two-sample t-test--Table 4}
        \item This paper lists all final (hyper-)parameters used for each model/algorithm in the paper’s experiments: \textbf{Yes, line numbers 303 to 306, and 350 to 351}.
    \end{itemize}
\end{itemize}

\clearpage
\onecolumn
\appendix
\section*{Appendix A: Description of Excel workbooks used for different code generation tasks}
\addcontentsline{toc}{section}{Appendix A: Description of Excel Workbooks Used for Different Code Generation Tasks}
\label{appendix: A}
\begin{table*}[!htbp] 
\caption{The distinct 28 workbook descriptions}
\label{tab: data_workbooks}
\centering
\scalebox{1}{
    \begin{tabular}{|p{0.15\textwidth}|p{0.7\textwidth}|p{0.045\textwidth}|}
    \hline
    \textbf{Work} &
    \textbf{Description} &
    \textbf{Count} \\ 
    
    \hline
    Boomerang Sales &
      Two tables: "Sheet1" for boomerang sales data and "Retail Price" for product prices. &
      9 \\ \hline
    Demographic Profile &
      Records a survey/questionaire respondent's demographic information. &
      7 \\ \hline
    Dragging &
      Contains data from an experiment where a hanging block (m2) drags another block (m1 = 0.75 kg) across a frictionless table via a rope connected to a frictionless, massless pulley. &
      8 \\ \hline
    Easy GDP Breakdown &
      Records economic indicators of various countries over multiple years. &
      10 \\ \hline
    Entire Shipping Costs &
      Records distances between a company’s customers and four delivery destinations. The per-mile shipping charge is \$3.5, with a minimum charge of \$80. &
      16 \\ \hline
    Entire Summer Sales &
      Captures a company’s sales data during the summer season. &
      13 \\ \hline
    Expense Report &
      Includes Tax and Total calculations: Tax = Subtotal * Tax rate; Total = Subtotal + Tax. &
      6 \\ \hline
    Future Value &
      Records multiple investments for which future values need to be calculated. Future Value = Present Value * (1 + Annual Interest Rate / \# of Compound Periods) \textasciicircum (Years * \# of Compound Periods). &
      7 \\ \hline
    GDP Breakdown &
      Contains two sheets: "Sheet1" records economic indicators across years for various countries, while "Sheet2" contains a list of selected country names. &
      7 \\ \hline
    Income Statement &
      Records a company's annual accounting data: Gross Profit = Net Sales - COGS; Operating Profit = Gross Profit - Operating Expenses; Net Profit = Operating Profit - Tax. &
      5 \\ \hline
    Income Statement 2 &
      Records a company's yearly accounting data. Gross Profit = Net Sales - Cost of Goods Sold (COGS); Net Sales = Sales - Sales Return - Discounts and Allowances; COGS = Material Charges + Labor Charges + Overhead. &
      9 \\ \hline
    Invoices &
      Records various invoices issued on different dates. &
      16 \\ \hline
    Maturity Date &
      Contains a list of loans and their respective lengths in days. &
      8 \\ \hline
    Net Income &
      Records a list of revenues and expenses. Net Income = Revenue - Total Expenses. &
      3 \\ \hline
    Period Rate &
      Logs annual rates for investments, considering that a year may contain multiple periods. Period Rate = Annual Rate / Periods per Year. &
      5 \\ \hline
    Present Value &
      Records investments requiring present value calculations. Present Value = Future Value / (1 + Annual Interest Rate / \# of Compound Periods) \textasciicircum (Years * \# of Compound Periods). &
      6 \\ \hline
    Pricing Table &
      Has two tables: "Sheet1," which contains transaction data (number of fence rolls sold on specific dates), and "Pricing Table," which determines the price per roll based on quantity ranges defined by "Units From" and "Unit To." &
      10 \\ \hline
    Ramp Up And Down &
      Records a block's acceleration in two scenarios but is incomplete. One scenario is documented in columns A to B, and the other in columns C to D. &
      5 \\ \hline
    Sales Rep &
      Captures monthly sales for each employee of a company. &
      6 \\ \hline
    Shipping Costs &
      Records distances between the company’s customers and four destinations. The per-mile shipping charge is \$3.11, with a minimum charge of \$75. &
      7 \\ \hline
    Simple Compound Interest &
      Calculates the interest on investments. Simple Interest = Principal Amount * Years * Interest Rate; Compound Interest = Principal Amount * (1 + Interest Rate) \textasciicircum Years - Principal Amount. &
      2 \\ \hline
    Small Balance Sheet &
      Records total assets, liabilities, and owners' equity. Assets = Current Assets + Fixed Assets + Other Assets; Liabilities \& Owner's Equity = Current Liabilities + Long Term Liabilities + Owner's Equity. &
      7 \\ \hline
    Stock Change &
      Tracks the values of a set of stocks on two separate dates. &
      4 \\ \hline
    Summer Sales &
      Records a company’s sales during the summer. &
      9 \\ \hline
    Tax &
      Logs a company's weekly sales, used to calculate taxes. Profit Before Tax = Sales - Total Expenses Before Tax; Tax Expense = Profit Before Tax * Tax Rate. &
      6 \\ \hline
    Velocity Displacement &
      Records velocity against displacement. &
      7 \\ \hline
    Weekly Sales &
      Records weekly sales and COGS data. Profit = Sales - COGS. &
      13 \\ \hline
    XY Scatter Plot &
      Shows how two variables, Range and Height, vary with the projection angle. &
      10 \\ \hline
    \textbf{Total} &
       &
      \textbf{221} \\ \hline
    \end{tabular}
}

\end{table*}

\newpage
\section*{Appendix B: Few-Shot Learning Experiment}\label{appendix: B}

We varied the number of provided few-shot examples from 1 to 19 to analyze the impact of few-shot learning on code-to-natural language translation (i.e., xwAPI to text) performance. We evaluated the resulting outputs using BLEU, GLEU, ROUGE-L, and METEOR on 20 test task instances. For this experiment, we selected GPT-4o-mini due to its cost efficiency and relatively large context window, enabling more extensive in-context learning. Then, we computed the mean scores for all four evaluation metrics and their corresponding margin of error with 95\% confidence interval using Equations \ref{equation: mean_metric} and \ref{equation: margin_of_error}, respectively.

\begin{figure*}[!hbp]
    \centering
    \includegraphics[width=1\linewidth]{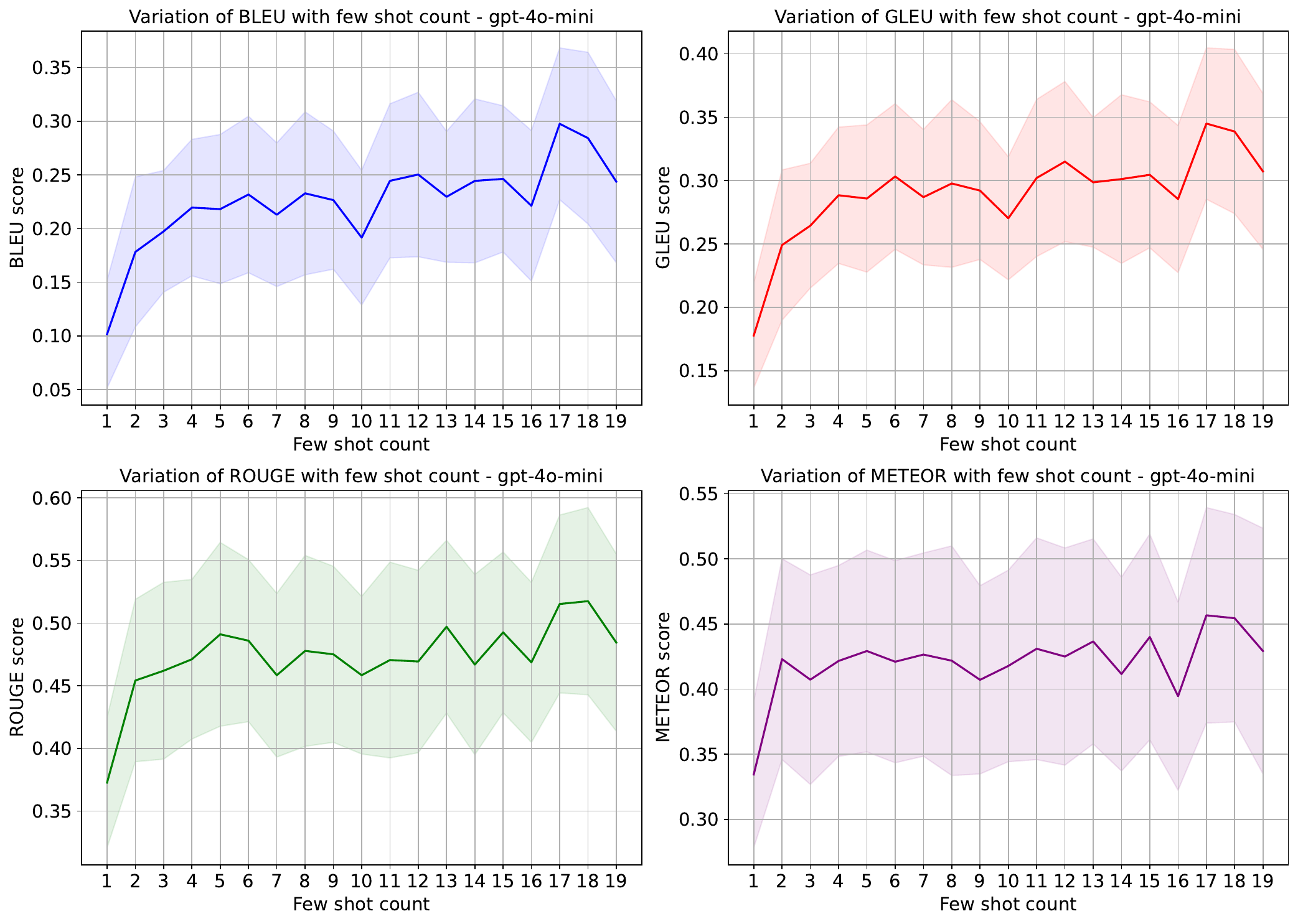}
    \caption{Variation of evaluation metrics with the number of few-shot examples}
    \label{fig: few_shot_experiment}
\end{figure*}

Figure \ref{fig: few_shot_experiment} illustrates the variation in evaluation metrics as a function of the number of few-shot examples. The figure shows performance across all four metrics, BLEU, GLEU, ROUGE-L, and METEOR plateaus after 4-shot learning, indicating diminishing returns beyond this threshold. Consequently, we adopted a 4-shot learning strategy for all LLMs except for Mixtral-8x7B. While using the Groq API, we identified that Mixtral-8x7B has a significantly smaller context window than other models. Due to this technical constraint, we employed a 1-shot learning configuration exclusively for Mixtral-8x7B to ensure compatibility with its limited context length.

\newpage
\section*{Appendix C: Representative example of a prompt used for code translation} \label{appendix: C}
The following listing presents the optimized prompt, refined using a prompt crafting framework for code translation.
\lstset{
    basicstyle=\small\ttfamily,
    columns=flexible,
    breaklines=true,
}
\begin{tcolorbox}[
    colback=gray!10,
    colframe=black,
    boxrule=1pt,
    arc=3mm,
    width=\textwidth
]
    \label{listing: code_translation_prompt}
    \begin{lstlisting}[caption={Code translation prompt used for GPT-4o-mini. For readability, portions of the content have been truncated and replaced with ellipses (...) to indicate omitted sections.},captionpos=b]
        {'role': 'developer', 'content': 'Summarize the each sub-step of instructions into explanations in natural language.
        Be brief and do not provide verbose explanations.
        Do not add text formatting such as bold, italic.
        Do not provide extra notes or postscriptum.
        Avoid redundant steps and provide minimal steps
        
        '}
        {'role': 'user', 'content': '- - Write(range="Sheet2!A1", value="Customers")
        - - Write(range="Sheet2!A1", value="Customers")
        - - CreateSheet(name="Sheet2")
        - - CreateSheet(sheetName="Sheet2")
        - - Write(range="Sheet2!A1", value="Customers")
          - Write(range="Sheet2!B1", value="Seattle")
          - Write(range="Sheet2!C1", value="Milwaukee")
          - Write(range="Sheet2!D1", value="Birmingham")
          - Write(range="Sheet2!E1", value="Oakland")
        .
          - Write(range="Sheet2!D10", value="=Sheet1!D10")
          - Write(range="Sheet2!E10", value="=Sheet1!E10")
        
        Here is the supplementary documentation you can reference:
        ['AutoFill(source: str, destination: str)...']
        Here is the corresponding sheet state:
        Sheet state: Sheet "Sheet1" has 5 columns (Headers are A: "Customers", B: "Seattle", C: "Milwaukee", D: "Birmingham", E: "Oakland") and 71 rows (1 header row and 70 data rows).
        
        {'role': 'assistant', 'content': '- Step 1. Write the headers in Sheet2.
        - Step 1. Create the headers in Sheet2.
        
        ...
        
        {'role': 'user', 'content': '- - Write(range="Sheet1!D1", value="Profit")
        - - Write(range="Sheet1!D2", value="=B2-C2")
        - - AutoFill(source="Sheet1!D2", destination="Sheet1!D2:D11")
        - - CreateChart(source="Sheet1!A1:D11", destSheet="Sheet1", chartType="Line", chartName="Weekly
            Trends")
        - - SetChartLegend(chartName="Weekly Trends", position="bottom")
        ...
        {'role': 'assistant', 'content': '- Step 1. Create a new column for Profit.
        - Step 2. Write the formula to calculate profit in the Profit column.
        - Step 3. Autofill the Profit formula for the remaining rows.
        - Step 4. Create a chart to show the trend of weekly Sales, COGS, and Profits.
        - Step 5. Set the legend for the chart at the bottom.
        
        {'role': 'user', 'content': '- - Filter(source="Sheet1!A1:E71", fieldIndex=3, criteria="<2000")
        ...
        Here is the corresponding sheet state:
        Sheet state: Sheet "Sheet1" has 5 columns (Headers are A: "Customers", B: "Seattle", C: "Milwaukee", D: "Birmingham", E: "Oakland") and 71 rows (1 header row and 70 data rows).
        ...
        
    \end{lstlisting}
\end{tcolorbox}

\newpage
\section*{Appendix D: JavaScript code generation from LLMs for spreadsheet manipulation using Retrieval-Augmented Generation (RAG)}
\label{appendix: D}
We attempted to generate JavaScript code for Excel manipulation using an LLM such as ChatGPT. However, the models frequently failed to produce error-free code due to a limited understanding of the Excel JavaScript API functions. Therefore, we implemented a Retrieval-Augmented Generation (RAG) pipeline using LangChain to provide the relevant API context.

\begin{figure}[!htbp]
    \centering
    \includegraphics[width=1\linewidth]{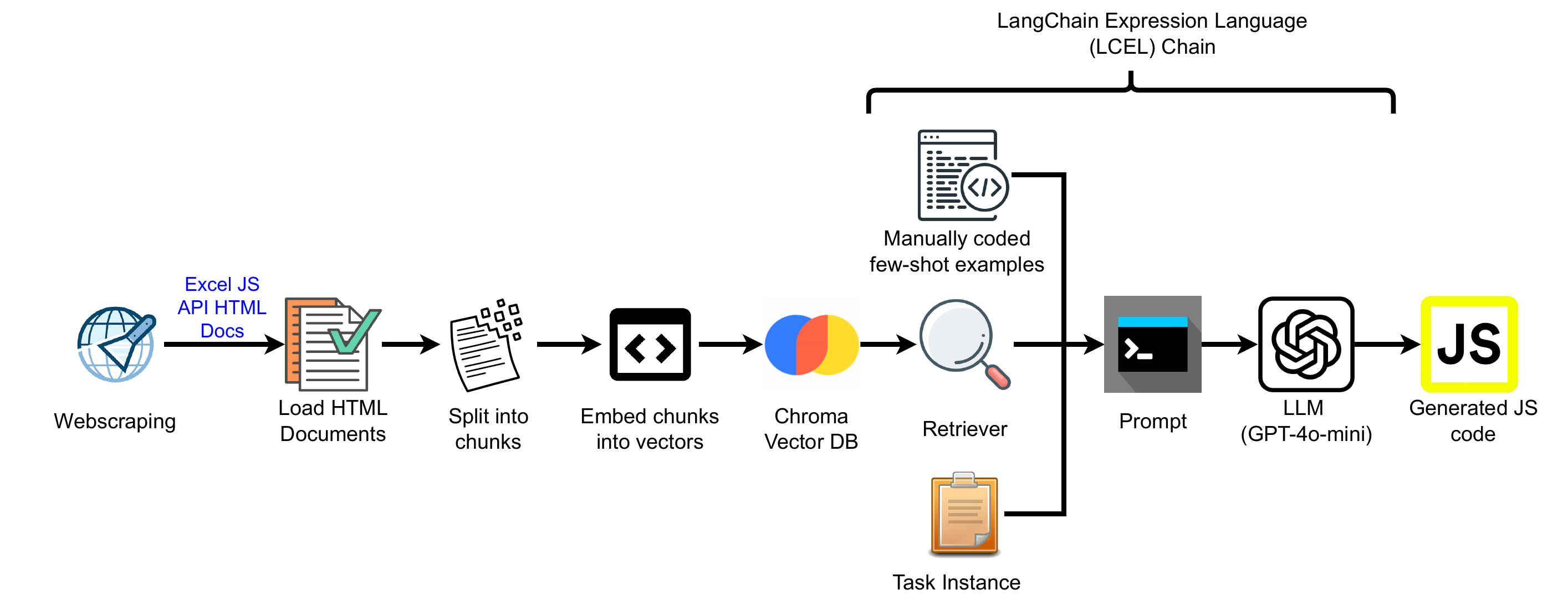}
    \caption{Retrieval Augmented Generation (RAG) pipeline used to generate Excel manipulation JavaScript code using LangChain}
    \label{fig: rag_pipeline}
\end{figure}

We started by web scraping the Microsoft Excel JavaScript API documentation to extract function references, as shown in Figure~\ref{fig: rag_pipeline}. We obtained 1129 HTML documents, segmented them into meaningful chunks using a semantic chunking strategy with the OpenAI text embedding model, and stored the resulting vectors in a Chroma database.

\begin{table}[!hbtp]
\centering
\caption{The evaluation results of the output from the RAG pipeline. Exec@1 and Pass@1 show the execution and pass rate, respectively.}

\label{tab: rag_results}
\renewcommand{\arraystretch}{1.5}
\begin{tabular}{|p{0.185\linewidth}|p{0.2\linewidth}|p{0.1\linewidth}|p{0.1\linewidth}|p{0.1\linewidth}|p{0.1\linewidth}|}
\hline
\diagbox[width=3.7cm]{Model}{Metric} & Total task instances count & Exec Count $\uparrow$              & Exec@1 $\uparrow$                  & Pass count $\uparrow$             & Pass@1 $\uparrow$                   \\ \hline
GPT-4o-mini  & \multicolumn{1}{r|}{20}    & \multicolumn{1}{r|}{12} & \multicolumn{1}{r|}{60\%} & \multicolumn{1}{r|}{4} & \multicolumn{1}{r|}{20\%} \\ \hline
\end{tabular}
\end{table}

To enable retrieval, we created a similarity-based retriever that selects the top 10 relevant chunks for a task. We also manually programmed two JavaScript code snippets as few-shot examples. During runtime, we inserted the benchmark task instance from SheetCopilot into the prompt as a runnable passthrough.

The final prompt for the LLM (GPT-4o-mini) consists of three parts: few-shot examples, context from a similarity search, and the task instance. This pipeline includes few-shot learning, vector retrieval, prompt construction, and LLM inference, implemented as a LangChain Expression Language (LCEL) chain.

While we implemented the RAG pipeline and JavaScript code generation, we did not automate the execution of the generated code in Excel. This automation is planned for future work. As a proof of concept, we randomly sampled 20 task instances from our dataset of 111 and generated JavaScript code using the RAG-enhanced GPT-4o mini pipeline. We then manually copied the code into Excel and executed it with the Script Lab plugin.

We evaluate results using metrics similar to those of SheetCopilot, specifically Exec@1 and Pass@1~\cite{li2024sheetcopilot}. Exec@1 measures the percentage of error-free executing JavaScript code, while Pass@1 assesses the code's functional correctness~\cite{chen2021evaluating}. Table~\ref{tab: rag_results} presents the results for 20 randomly selected samples.

Our RAG pipeline achieved an Exec@1 score of 60\% and a Pass@1 score of 20\%, highlighting its potential for automated JavaScript code generation in spreadsheet manipulation. Furthermore, the RAG pipeline can be improved through better fine-tuning, expanded few-shot configurations, and architectural refinements. However, these enhancements are beyond the scope of this study and are suggested for future work. Our RAG pipeline implementation details are available in our GitHub repository\footnote{https://github.com/AmilaIndika789/SheetCopilot}.

\section*{Appendix E: Comparison of pairwise two-sample t-test for Llama-3.3-70B, GPT-4o, and GPT-4o-mini}

\begin{table*}[!htbp]
\centering
\caption{Summary of two-sample t-test results, reporting t-values and p-values. Pairwise two-sample t-tests comparing Llama-3.3-70B, GPT-4o, and GPT-4o-mini show no statistically significant differences in evaluation metrics, with all p-values exceeding 0.05. The models perform equivalently on the SOD benchmark.}
\label{tab: two_sample_t_test}
\renewcommand{\arraystretch}{1.5}
\begin{tabular}{|l|rr|rr|rr|rr|}
\hline
\multirow{2}{*}{\diagbox[width=4cm]{Models}{Metric}} &
  \multicolumn{2}{c|}{BLEU} &
  \multicolumn{2}{c|}{GLEU} &
  \multicolumn{2}{c|}{ROUGE-L} &
  \multicolumn{2}{c|}{METEOR} \\ \cline{2-9} 
 &
  \multicolumn{1}{c|}{t-value} &
  \multicolumn{1}{c|}{p-value} &
  \multicolumn{1}{c|}{t-value} &
  \multicolumn{1}{c|}{p-value} &
  \multicolumn{1}{c|}{t-value} &
  \multicolumn{1}{c|}{p-value} &
  \multicolumn{1}{c|}{t-value} &
  \multicolumn{1}{c|}{p-value} \\ \hline
Llama-3.3, GPT-4o &
  \multicolumn{1}{r|}{0.122} &
  0.903 &
  \multicolumn{1}{r|}{0.243} &
  0.808 &
  \multicolumn{1}{r|}{0.363} &
  0.717 &
  \multicolumn{1}{r|}{-1.205} &
  0.228 \\ \hline
Llama-3.3, GPT-4o-mini &
  \multicolumn{1}{r|}{-0.317} &
  0.751 &
  \multicolumn{1}{r|}{-0.231} &
  0.817 &
  \multicolumn{1}{r|}{-0.346} &
  0.729 &
  \multicolumn{1}{r|}{-1.346} &
  0.178 \\ \hline
GPT-4o, GPT-4o-mini &
  \multicolumn{1}{r|}{-0.435} &
  0.664 &
  \multicolumn{1}{r|}{-0.469} &
  0.639 &
  \multicolumn{1}{r|}{-0.726} &
  0.468 &
  \multicolumn{1}{r|}{-0.156} &
  0.876 \\ \hline
\end{tabular}
\end{table*}

\end{document}